\documentclass[a4paper,conference]{IEEEtran}

\usepackage[T1]{fontenc}            
\usepackage[utf8]{inputenc}         
\usepackage{cite}
\usepackage{amsmath,amssymb,amsfonts}
\usepackage{algorithmic}
\usepackage{graphicx}
\usepackage{textcomp}
\usepackage{xcolor}
\usepackage{booktabs}
\usepackage{tabularx}
\usepackage{url}
\usepackage{dblfloatfix}
\usepackage{capt-of}
\usepackage[hidelinks]{hyperref}
\hypersetup{draft,bookmarks=false,pdfborder={0 0 0}} 
\usepackage{xurl}
\usepackage{microtype}
\usepackage{orcidlink} 
\graphicspath{{./}}

\def\BibTeX{{\rm B\kern-.05em{\sc i\kern-.025em b}\kern-.08em
    T\kern-.1667em\lower.7ex\hbox{E}\kern-.125emX}}

\usepackage{etoolbox}
\makeatletter
\@ifpackageloaded{hyperref}{\renewcommand{\href}[2]{#2}\let\url\nolinkurl}{}
\makeatother

\usepackage{fancyhdr}
\usepackage[hidelinks]{hyperref} 

\fancypagestyle{arxivfooter}{
    \fancyhf{} 
    \fancyfoot[C]{ 
        \begin{minipage}{\textwidth}
            \centering\scriptsize 
            \textbf{Published in:} 2025 IEEE Asia Pacific Conference on Wireless and Mobile (APWiMob), Bali, Indonesia, 2025, pp. 180-186. \\
            \textbf{DOI:} \href{https://doi.org/10.1109/APWiMob67231.2025.11269122}{10.1109/APWiMob67231.2025.11269122} \\
            \copyright 2025 IEEE. Personal use of this material is permitted. Permission from IEEE must be obtained for all other uses, in any current or future media, including reprinting/republishing this material for advertising or promotional purposes, creating new collective works, for resale or redistribution to servers or lists, or reuse of any copyrighted component of this work in other works.
        \end{minipage}
    }
}

\begin{document}

\title{Web Technologies Security in the AI Era: A Survey of CDN-Enhanced Defenses}



\author{%
\IEEEauthorblockN{
Mehrab Hosain\,\orcidlink{0009-0007-1079-1052},
Sabbir Alom Shuvo\,\orcidlink{0000-0002-7124-2459},
Matthew Ogbe\,\orcidlink{0009-0009-8753-4745},\\
Md Shah Jalal Mazumder\,\orcidlink{0009-0007-5554-408X},
and Yead Rahman\,\orcidlink{0009-0008-1742-1027}
}
\IEEEauthorblockA{
Louisiana Tech University, Ruston, LA, USA\\
Emails: robinhosain@gmail.com, shovon961@gmail.com, matthewogbe@gmail.com,\\
mma112@latech.edu, yead1409031@gmail.com
}

\vspace{0.5em} 

\IEEEauthorblockN{
Md Azizul Hakim\,\orcidlink{0000-0002-5129-2408},
and Anukul Pandey\,\orcidlink{0000-0003-2737-112X}
}
\IEEEauthorblockA{
Delhi Technological University, Delhi, India\\
Emails: ahanik94@gmail.com, anukul.pandey@dtu.ac.in
}
}

\maketitle
\thispagestyle{arxivfooter} 
\bstctlcite{IEEEexample:BSTcontrol,nodash}

\begin{abstract}

The modern web stack, which is dominated by browser-based applications and API-first backends, now operates under an adversarial equilibrium where automated, AI-assisted attacks evolve continuously. Content Delivery Networks (CDNs) and edge computing place programmable defenses closest to users and bots, making them natural enforcement points for machine-learning (ML) driven inspection, throttling, and isolation. This survey synthesizes the landscape of AI-enhanced defenses deployed at the edge: (i) anomaly- and behavior-based Web Application Firewalls (WAFs) within broader Web Application and API Protection (WAAP), (ii) adaptive DDoS detection and mitigation, (iii) bot management that resists human-mimicry, and (iv) API discovery, positive security modeling, and encrypted-traffic anomaly analysis. We add a systematic survey method, a threat taxonomy mapped to edge-observable signals, evaluation metrics, deployment playbooks, and governance guidance. We conclude with a research agenda spanning XAI, adversarial robustness, and autonomous multi-agent defense. Our findings indicate that edge-centric AI measurably improves time-to-detect and time-to-mitigate while reducing data movement and enhancing compliance, yet introduces new risks around model abuse, poisoning, and governance.
\end{abstract}


\section{Introduction}
The web's attack surface has expanded with microservices, multi-tenant APIs, and client-heavy single-page apps. Loss magnitudes and incident volumes continue to rise: independent analyses report record breach costs and sustained pressure on security teams \cite{ref1,ref4}. Attacker automation amplifies velocity and scale: WAAP telemetry shows sharp increases in web and API-layer attacks (credential stuffing, injection variants, and bot traffic) \cite{ref4}. These dynamics compel a shift from reactive signatures to adaptive, data-driven controls that learn normality and spotlight deviations \cite{ref4,ref5}.

CDNs and edge platforms have matured from content accelerators to programmable security planes. By inspecting and acting on flows at Points of Presence (PoPs), they shave detection latency and offload origin infrastructure \cite{ref6,ref18}. With integrated WAF/WAAP, bot controls, and API gateways \cite{ref24,ref25,ref30,ref31,ref32,ref34}, the edge becomes an effective, privacy-aware enforcement and learning layer \cite{ref19,ref20}. 

\noindent\textbf{Why now?} Three macro-shifts make AI-at-the-edge decisive. First, attacker \emph{capability} has scaled: turnkey bot services, residential proxies, and LLM-aided payload generation compress the time between disclosure and weaponization \cite{ref4,ref11}. Second, defender \emph{visibility} has migrated to CDNs and API gateways, where most requests are seen first and where enforcement is cheapest \cite{ref24,ref25,ref28}. Third, \emph{regulatory} pressure (GDPR/AI Act) demands privacy-preserving data flows, favoring learning and inference close to the user \cite{ref19,ref20}.

\noindent\textbf{Scope and exclusions.} We focus on web and API threats observable at HTTP(S)/QUIC layers, and on mitigations operable from PoPs. We exclude email/DNS-only threats and purely endpoint EDR controls unless they integrate with CDN/WAAP pipelines.

\noindent\textbf{Position in literature.} Prior surveys broadly cover AI in cybersecurity \cite{ref3,ref9} or DDoS/edge learning \cite{ref7,ref6}; our contribution centers on \emph{CDN-integrated WAAP}, threat--signal mappings unique to the edge, and deployable playbooks with SLOs, privacy techniques, and platform comparisons \cite{ref24,ref25,ref30,ref31,ref32,ref34}.

\noindent\textbf{Contributions.} We (1) provide a methodical survey of AI/ML defenses in CDN contexts; (2) present a threat taxonomy and edge-signal mapping; (3) detail model/feature design patterns; (4) propose deployment playbooks with SLO-centric evaluation; (5) extend platform comparisons; and (6) outline governance and future directions.


\subsection{Survey Methodology}
We performed a structured survey focused on edge/CDN-deployable web and API defenses. Sources: ACM Digital Library, IEEE Xplore, arXiv; plus transport standards and public datasets (TLS 1.3, QUIC \cite{ref64,ref63}; CSE-CIC-IDS2018, UNSW-NB15 \cite{ref67,ref65}). Inclusion: 2020--2025 works on AI/ML for L7 DDoS, bots, API abuse, or WAAP under HTTP(S)/QUIC. Exclusion: non-web protocols, email/DNS-only defenses, and endpoint-only EDR without a CDN/PoP enforcement path. For each study we extracted attack class, edge-feasible signals/features (incl. encrypted-traffic side-channels), model family, evaluation metrics, edge/PoP placement, and privacy/governance technique, then normalized actions to an operational ladder (observe $\rightarrow$ shape $\rightarrow$ challenge $\rightarrow$ block) and synthesized SLO-aware playbooks and a vendor capability map.   \noindent\textit{Limitations:} heterogeneous datasets and mixed academic/practitioner evidence.

\section{Foundations and threat Model}
\subsubsection{Common Web and API Weaknesses}
Applications remain exposed to injection, XSS, and CSRF; APIs add mass assignment and broken object-level authorization. OWASP rankings highlight these as enduring, high-impact risks \cite{ref61,ref62}. DDoS continues to degrade availability through L3/4 floods and stealthy L7 bursts \cite{ref7,ref9}. A significant fraction of API incidents occur \emph{after} authentication \cite{ref4}. Sophisticated bots emulate cursor/timing and device fingerprints, complicating differentiation \cite{ref13,ref31}.

\subsubsection{AI/ML Building Blocks}
Supervised, semi-supervised, and unsupervised learning drive detection and triage; deep models (CNN/RNN) capture higher-order traffic/sequence structure for IDS and DDoS \cite{ref9}. Reinforcement learning (RL) adapts throttling/routing online \cite{ref7}. Large Language Models (LLMs) assist rule authoring and fuzzing but are dual-use \cite{ref11,ref58}.

\subsubsection{Notation and Anomaly Scoring}
Let $\mathbf{x}\in\mathbb{R}^d$ denote request features (method, header lengths, timing) with baseline distribution $\mathcal{D}_0$. An anomaly score $s(\mathbf{x})$ via density model $p_\theta(\mathbf{x})$ or distance-to-manifold is
\begin{equation}
s(\mathbf{x})=-\log p_\theta(\mathbf{x})\quad\text{or}\quad s(\mathbf{x})=\|\mathbf{x}-\Pi_{\mathcal{M}}(\mathbf{x})\|_2.
\end{equation}
A request is anomalous if $s(\mathbf{x})>\tau$, where $\tau$ is tuned to meet a false-positive budget (FPB).

\subsubsection{System Model and Assumptions}
Anycast CDN PoPs with programmable request flows (workers/functions), per-request features (headers, timing, TLS/QUIC metadata), and per-session linkage (cookies, tokens) \cite{ref24,ref28}. Origin servers are shielded behind the CDN; API gateways enforce identity and shape traffic. TLS~1.3/QUIC are preferred \cite{ref64,ref63}.

\noindent\textbf{Adversary model:} Adversaries may (i) control distributed botnets and rotating egress IPs/ASNs; (ii) mimic user behavior at micro-timing resolution; (iii) obtain valid credentials or sessions; (iv) exploit zero-days at the app/API layer; and (v) probe rate and policy edges. We assume no compromise of CDN infrastructure or private keys.

\noindent\textbf{Defender objectives.} Minimize attack containment time (ACT), false-positive rate (FPR) on critical flows (login/checkout), and added P95/P99 latency, subject to privacy constraints \cite{ref9}. 

\subsubsection{Identity \& Session Realities}
API ecosystems rely on OAuth/OIDC tokens and service identities; typical pitfalls include excessive token lifetimes, unscoped permissions, and predictable refresh patterns. Post-auth abuse (A2 in OWASP API) motivates \emph{behavioral} and \emph{sequence}-aware detection at the edge \cite{ref61,ref62}.

\subsection{AI/ML-Powered Defenses}
Signature-only WAFs miss novel and polymorphic payloads. AI-enabled controls add anomaly detection, behavioral analytics, and generative augmentation to harden rules \cite{ref9,ref11,ref24,ref30}. \emph{In practice}, WAAP suites fuse CDN, DNS, WAF, API gateways, and bot protection with centralized policy and telemetry \cite{ref24,ref25,ref31,ref32,ref34}.

\noindent\textbf{Pipelines and feedback:} Practical stacks couple rules and ML: fast signatures for known payloads, anomaly scores for novelty, and \emph{triage actions} (observe/shape/challenge/block). Labels flow back from SOC adjudication to the feature store for continuous retraining \cite{ref5,ref22}. Confidence calibration (e.g., Platt scaling) stabilizes thresholds across regions and traffic seasons.

\noindent\textbf{Mitigation semantics:} Actions are graduated: (i) \emph{observe} (shadow scoring only), (ii) \emph{shape} (rate-limit or delay), (iii) \emph{challenge} (turnstiles, step-up auth), and (iv) \emph{block}. For login and payments, we bound FPR with explicit budgets and prefer challenge over block until multi-signal consensus is reached.

\noindent\textbf{Model efficiency:} At PoPs, distillation and quantization shrink deep models; shallow ensembles (GBDT + sequence GRU) often meet latency targets while preserving recall. Models are versioned, canaried, and rolled back per-region under policy-as-code.

\subsubsection{DDoS Detection and Mitigation}
\begin{figure}[!t]
\centerline{\includegraphics[width=0.8\columnwidth]{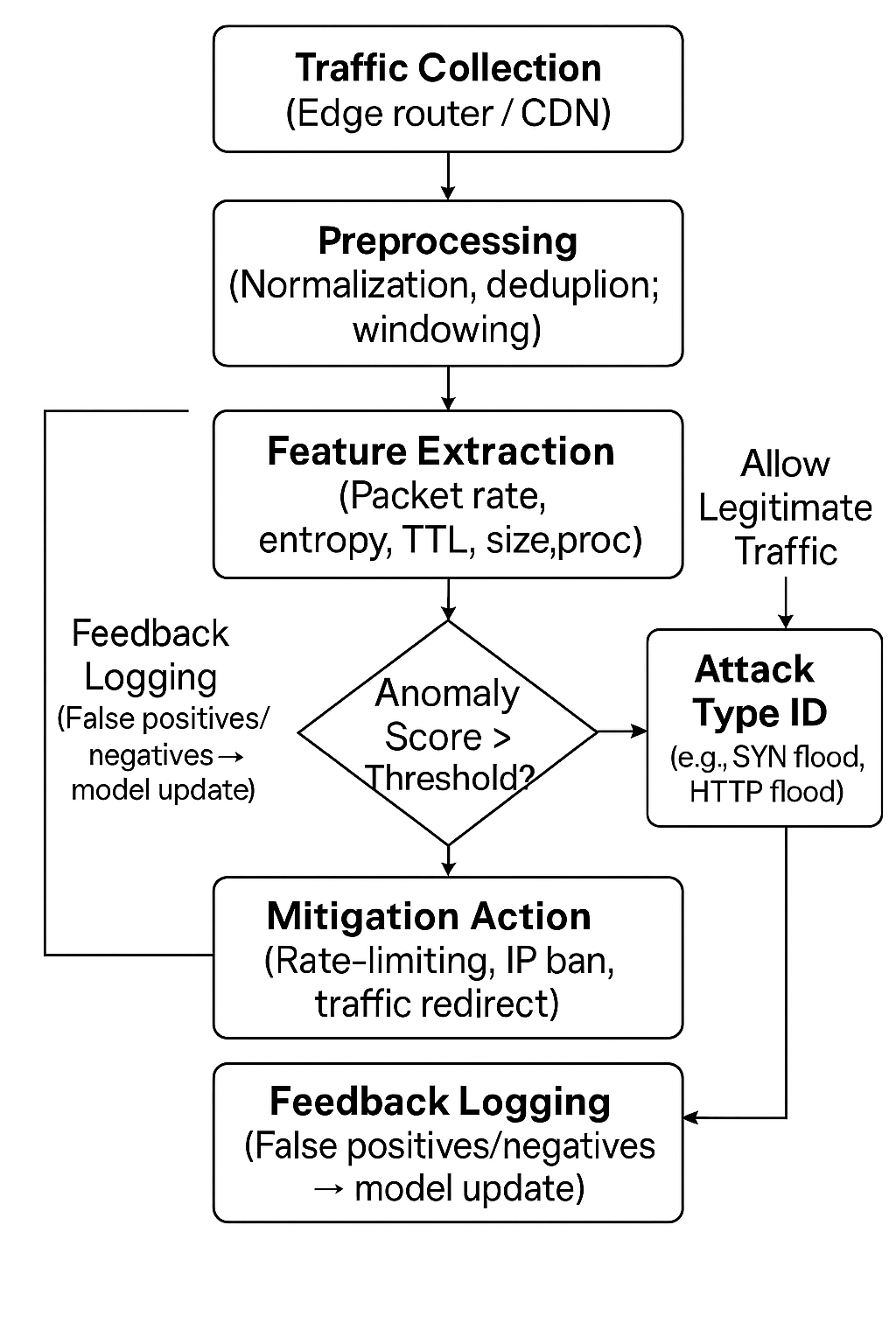}}
\caption{AI/ML DDoS pipeline from edge telemetry to adaptive mitigation with feedback.}
\label{fig:ddos_pipeline}
\end{figure}

ML/DL classifiers label flows using rate, entropy, header consistency, and temporal features; deep models capture subtle periodicity and cross-feature interactions \cite{ref7,ref9}. DRL policies adapt rate limits, challenge puzzles, or upstream blackholing with minimal collateral impact \cite{ref7}. \emph{Volumetric vs. application-layer:} L3/4 floods are absorbed by Anycast and scrubbing; L7 attacks target origin-expensive endpoints (search, invoice export) with cache-bypass headers. Edge classifiers prioritize \emph{cost-aware} features (cache status, origin timeouts) and preemptively shape traffic \cite{ref7,ref24}. Figure~\ref{fig:ddos_pipeline} illustrates the pipeline.

\subsubsection{Bot Management and Human-Mimicry Resistance}
\begin{figure}[!t]
\centerline{\includegraphics[width=0.9\columnwidth]{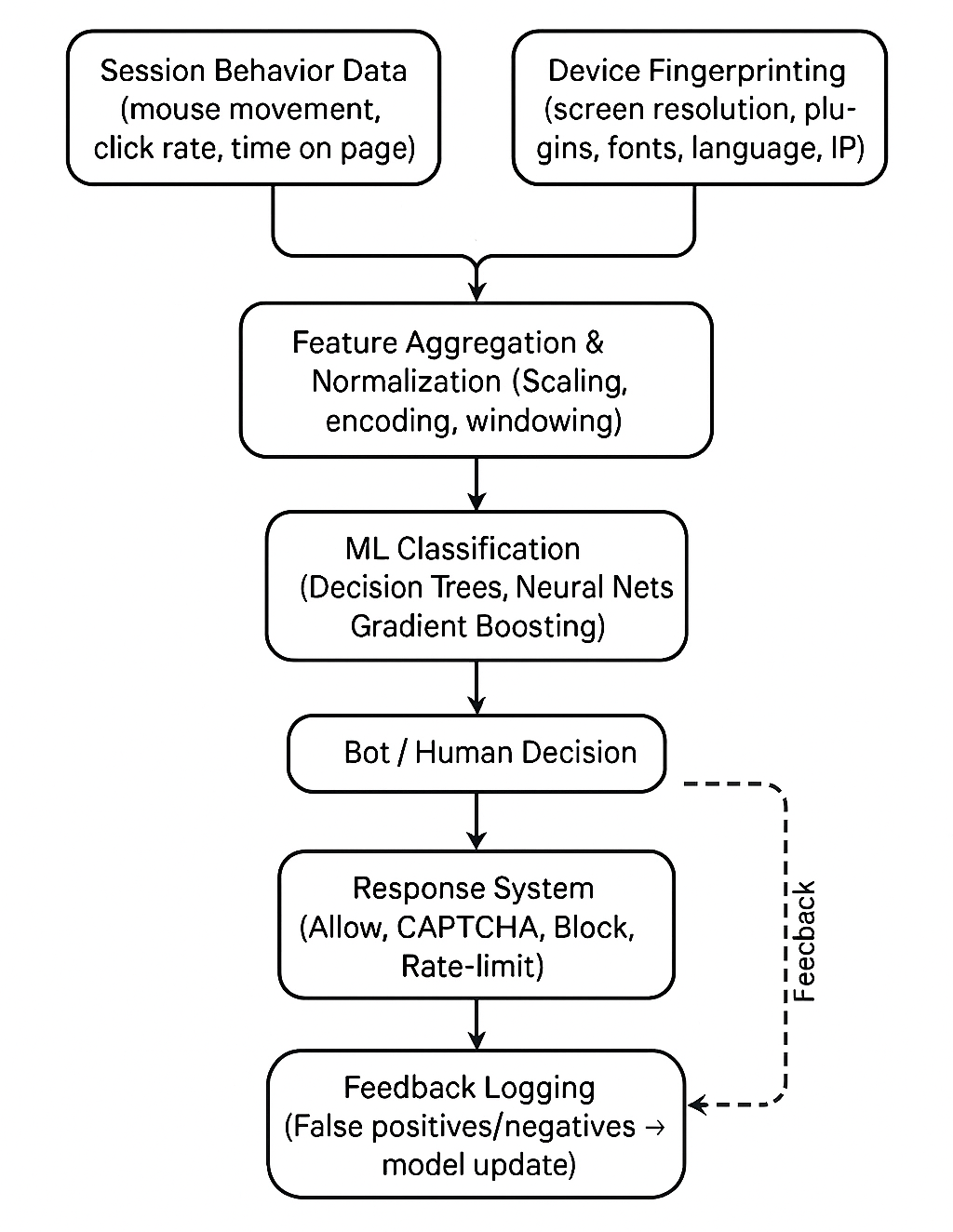}}
\caption{Bot detection flow: behavior + fingerprint features $\rightarrow$ ML classification $\rightarrow$ allow/challenge/block with continuous learning.}
\label{fig:bot_pipeline}
\end{figure}

Effective defenses correlate micro-movements, dwell times, layout shifts, WebGL/Canvas traits, TLS ClientHello, and history consistency. Models include gradient-boosted trees and deep sequence encoders; privacy-aware feature selection is critical under GDPR/AI Act \cite{ref13,ref31,ref20}. Providers report sizable YoY growth in blocked malicious bot traffic \cite{ref4}. Figure~\ref{fig:bot_pipeline} shows the flow.

\subsubsection{API Security}
Needs include: (i) automated discovery \& inventory, (ii) positive security modeling, (iii) encrypted-traffic anomaly detection \cite{ref24,ref28,ref32}.
\begin{itemize}
\item \textbf{Discovery.} Edge observation + service-mesh catalogs reduce ``shadow APIs'' \cite{ref24,ref32}.
\item \textbf{Positive models.} Learned verbs, parameter shapes, and rate envelopes block unexpected calls \cite{ref24,ref30}.
\item \textbf{Encrypted anomaly detection.} Side-channel features and sequences provide DPI-free detection over TLS/QUIC \cite{ref28,ref63,ref64}.
\end{itemize}

\section{Threat Taxonomy and Edge-Signal Mapping}
Table~\ref{tab:taxonomy} maps common threats to edge-observable signals, ML tasks, and recommended actions.

\begin{table}[!t]
\caption{Expanded Threat $\rightarrow$ Edge Signals $\rightarrow$ ML Task $\rightarrow$ Action}
\label{tab:taxonomy}
\centering
\begin{tabularx}{\columnwidth}{@{}lX@{}}
\toprule
\textbf{Threat} & \textbf{Signals / Task / Action} \\ \midrule
SQLi/XSS & Param entropy spikes; rare method+path pairs; encoder mismatches $\rightarrow$ anomaly + supervised WAF $\rightarrow$ normalize/block \cite{ref9,ref30}.\\
Credential Stuffing & POST/login bursts; stable JA3/UA with rotating IPs; low dwell $\rightarrow$ seq. model + risk score $\rightarrow$ challenge/shape \cite{ref24,ref25}.\\
Account Takeover (ATO) & Geo/ASN drift post-login; device mismatch; password reset storms $\rightarrow$ session anomaly $\rightarrow$ step-up auth \cite{ref61}.\\
L7 DDoS & Cache-bypass; synchronized bursts; low payload entropy $\rightarrow$ DL classifier $\rightarrow$ shape/blackhole \cite{ref7}.\\
API Abuse/Discovery & Verb/schema mismatch; high 4xx on hidden routes $\rightarrow$ positive model $\rightarrow$ block \cite{ref24,ref32}.\\
Scraping Bots & Absent micromotions; periodic timing; headless traits $\rightarrow$ ensemble classifier $\rightarrow$ tarpit/challenge \cite{ref13,ref31}.\\
SSRF Probing & Unusual callback targets; metadata-service patterns $\rightarrow$ rule+anomaly $\rightarrow$ deny egress \cite{ref62}.\\
Payment Abuse & Velocity on checkout; BIN concentration; mismatch signals $\rightarrow$ supervised fraud model $\rightarrow$ step-up/block.\\ \bottomrule
\end{tabularx}
\end{table}

\subsection{Edge Feature and Model Patterns}
Table~\ref{tab:features} summarizes robust PoP-level features (transport/TLS, HTTP semantics, timing, behavioral, fingerprint, sequences, geo/ASN).

\begin{table}[!t]
\caption{Edge Feature Catalog for WAAP Inference (non-exhaustive)}
\label{tab:features}
\centering
\begin{tabularx}{\columnwidth}{@{}lX@{}}
\toprule
\textbf{Category} & \textbf{Examples at PoP} \\ \midrule
Transport/TLS & SNI presence, JA3/JA4 hash, RTT, handshake retries, QUIC spin-bit \cite{ref63}. \\
HTTP Semantics & Method mix, header length distro, referrer anomalies. \\
Timing/Rate & Inter-arrival deltas, burstiness, diurnal alignment (EWMA/CUSUM). \\
Behavioral & Dwell time, click/scroll cadence, cursor micromotions. \\
Fingerprint & Canvas/WebGL, fonts, media queries, plugin vector. \\
Sequence & Login$\rightarrow$enum$\rightarrow$reset, API verb chains. \\
Geo/ASN & IP reputation, ASN stability, exit-node heuristics. \\ \bottomrule
\end{tabularx}
\end{table}

\subsubsection{Streaming Detection Primitives}
Using an exponentially weighted moving average (EWMA) on rate $r_t$:
\begin{equation}
m_t=\alpha r_t+(1-\alpha)m_{t-1}, \quad z_t=\frac{r_t-m_t}{\sigma_t},
\end{equation}
raise an alert when $z_t>\gamma$. Coupling with CUSUM reduces drift sensitivity. Thresholds $(\gamma,\alpha)$ are tuned to an FPB on critical endpoints.

\subsubsection{Concept Drift \& Personalization}
Traffic mixes vary by region, time, and campaign season. Drift monitors (PSI/JS divergence) trigger retraining or per-region thresholds. Personalization (\emph{federated fine-tuning}) adapts models to regional patterns while keeping global weights stable \cite{ref19}.

\subsubsection{Calibration and Ensembling}
Score calibration improves action stability across PoPs; stacking (GBDT on static features + RNN on sequences) reduces variance. For encrypted traffic, side-channel features (sizes, timings, TLS/QUIC metadata) remain reliable under TLS~1.3/QUIC \cite{ref64,ref63}.

\subsubsection{Interplay with Caching}
Security and performance co-design: avoid cache-key pollution by normalizing parameters pre-classification; prefer shaping before origin fetch; track \emph{cache-HIT delta} as a first-class SLO during mitigations.

\begin{table}[!t]
\caption{Operational Differentiators at the Edge}
\label{tab:opsdiff}
\centering
\begin{tabularx}{\columnwidth}{@{}lX@{}}
\toprule
\textbf{Dimension} & \textbf{What to Measure} \\ \midrule
Policy Propagation & Time from rule commit to global enforcement. \\
Regionalization & Ability to vary thresholds/models per jurisdiction. \\
Telemetry Export & Real-time streaming to SIEM/LS with privacy filters. \\
Canary Controls & Per-PoP rollouts and auto-rollback. \\
Label Loop & SOC adjudications $\rightarrow$ feature store latency.\\ \bottomrule
\end{tabularx}
\end{table}

\subsection{Why the Edge Matters}
\begin{figure}[!t]
\centerline{\includegraphics[width=0.8\columnwidth]{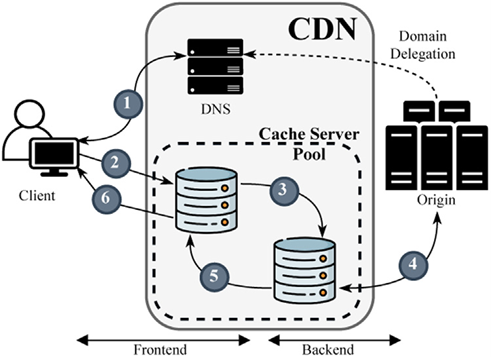}}
\caption{AI-enhanced CDN path: classification and policy decisions at PoPs prior to origin contact.}
\label{fig:cdn_arch}
\end{figure}

Anycast steers users to nearest PoPs, reducing latency and egress to origin. Edge inference curbs backhaul of sensitive data and enables regional policy variation \cite{ref6,ref18,ref24}. Figure~\ref{fig:cdn_arch} highlights the security path.

\subsubsection{Privacy-Preserving Learning at PoPs}
\begin{figure}[!t]
  \centering
  \includegraphics[width=0.9\columnwidth]{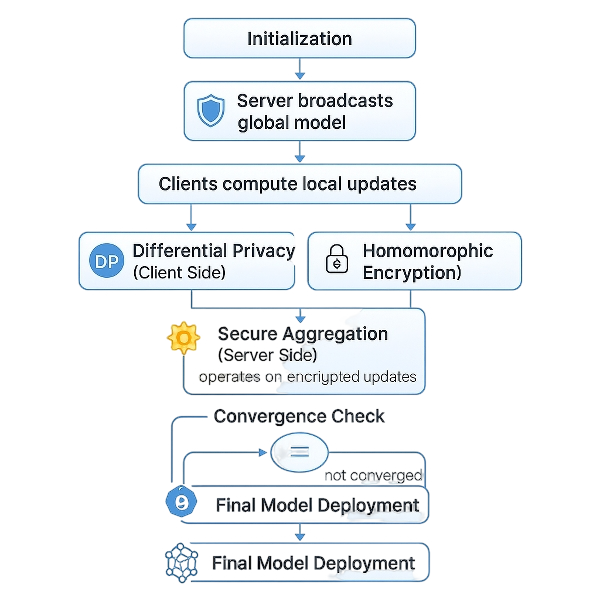}
  \caption{Federated learning with DP/HE for privacy-preserving edge training and global aggregation.}
  \label{fig:federated}
\end{figure}

Federated learning (FL) trains models across PoPs/devices using local gradients; secure aggregation and differential privacy (DP) bound leakage; homomorphic encryption (HE) can protect updates \cite{ref19}. This aligns with GDPR/AI Act guardrails \cite{ref20}. Figure~\ref{fig:federated} sketches the approach.

\begin{table*}[t]
\caption{Leading CDN/WAAP Platforms and Security Capabilities }
\label{tab:waap_large}
\footnotesize
\begin{tabularx}{\textwidth}{@{}p{0.14\textwidth} p{0.14\textwidth} p{0.33\textwidth} >{\hspace{0.5em}}p{0.33\textwidth}@{}}
\toprule
\textbf{Platform} & \textbf{Core CDN/Edge Strengths} & \textbf{AI/ML-Enhanced Security Features} & \textbf{Notable Differentiators / Notes} \\
\midrule
Cloudflare \cite{ref24} &
Anycast, large PoP footprint &
WAF w/ ML detection; adaptive DDoS heuristics+ML; behavioral bot mgmt; API discovery &
Integrated DNS/CDN/WAF/Zero Trust; global policy + telemetry. \\
\addlinespace
AWS CloudFront \cite{ref25,ref27,ref26} &
Global AWS integration &
AWS WAF + AWS Shield; Bot Control w/ predictive ML; per-resource policies &
IAM alignment; mature compliance posture; native to AWS stack. \\
\addlinespace
Google Cloud CDN \cite{ref28,ref29} &
GCP + global load balancing &
Cloud Armor adaptive protection; sensitive-data tooling; AI posture guidance &
Tight LB/VPC firewall integration; Google threat intel. \\
\addlinespace
Akamai \cite{ref30,ref31,ref32} &
Massive edge + threat intel &
Self-learning WAF; deep-learning-driven bot detection; API PII identification &
Enterprise controls; strong bot and API discovery features. \\
\addlinespace
Azure CDN \cite{ref34} &
Microsoft stack integration &
WAF anomaly ML; ML-tuned DDoS; “AI gateway” for API mgmt &
Enterprise observability; policy via Azure ecosystem. \\
\addlinespace
Fastly (Next-Gen WAF) \cite{refFastlyProd,refFastlyDocs,refFastlyDS} &
High-perf edge; developer workflow focus &
Signal Sciences--powered WAAP; behavioral signals; advanced rate limiting; ML-assisted detections &
Hybrid deployments (edge/agent); strong DevOps integrations and “signals” model. \\
\addlinespace
Imperva WAAP \cite{refImpervaAppSec,refImpervaWAAP,refImpervaWAF} &
Global cloud POPs; CDN add-ons &
Unified WAAP: WAF, DDoS, Bot Mitigation, API Security; ML/analytics across modules &
Single-console operations; PCI DSS 4.0 support materials; WAAP buyer guides. \\
\addlinespace
F5 Distributed Cloud WAAP \cite{refF5Soln,refF5WAF,refF5Gloss} &
Volterra-based distributed cloud &
F5 WAF + Shape Bot Defense; automatic signature tuning; behavior engine; AI-assisted policies &
Hybrid/multi-cloud posture; “virtual patching” + policy-as-code across estates. \\
\addlinespace
Radware (Cloud) \cite{refRadwareWebDDoS,refRadwareDPA,refRadwareAWS} &
Global scrubbing network &
Behavioral L7 (Web DDoS/Tsunami) detection+mitigation; encrypted L7 handling; bot mgr options &
Focus on sophisticated HTTP/S floods; network behavioral analytics. \\
\addlinespace
CDNetworks WAAP \cite{refCDN-WAAP-Brochure,refCDN-WAAP-Checklist} &
APAC-strong CDN; performance tuning &
AI-assisted WAAP incl.\ API discovery; ML-based bot mgmt; integrated DDoS/WAF/API protection &
Full-lifecycle API discovery; flood shield options; global + regional policies. \\
\addlinespace
Azion (Edge Firewall/WAAP) \cite{refAzionWAAP,refAzionDocsWAF,refAzionEdgeFW} &
Programmable edge firewall/rules engine &
WAF + DDoS + bot protections; rules-engine automation; SIEM streaming; TLS/QUIC aware &
Strong developer control (Terraform/APIs); 100\% SLA claims for network. \\
\addlinespace
Bunny.net (Bunny Shield) \cite{refBunnyShield,refBunnyBlog,refBunnyDocs} &
Cost-efficient global CDN &
“Next-Gen AI WAF”, DDoS, global rate limiting, bot detection; API integration for config &
Rapidly evolving (2025): complex bot detection; API protection roadmap maturing. \\
\addlinespace
Cloudbric \cite{refCloudbricWAF,refCloudbricCDN} &
CDN + managed WAF bundle &
Managed WAF with DDoS protection; customizable rules; SSL; SMB-friendly packaging &
Proxy-style deployment; emphasis on simplicity for smaller teams. \\
\addlinespace
Edgio (WAAP\textbf{)} \cite{refEdgioDDoS,refEdgioCSPress} &
Edge network + app platform (status changed in 2025) &
WAAP incl.\ advanced rule customizer, client-side protection; bot and DLP features reported &
\emph{Note:} Corporate update (Aug 2025): Chapter 11; Akamai acquired select assets—-offerings may change regionally. \\
\bottomrule
\end{tabularx}
\end{table*}
\section{Deployment Patterns and Evaluation}
\subsection{Deployment Patterns and Playbooks}
\subsubsection{Decision Ladder at the Edge}
\textit{observe} (shadow only) $\rightarrow$ \textit{score} (attach $s(\mathbf{x})$) $\rightarrow$ \textit{shape} (rate-limit) $\rightarrow$ \textit{challenge} (turnstile/step-up) $\rightarrow$ \textit{block}.

\subsubsection{Safe Rollouts and MLOps}
Loop: (1) train from PoP telemetry + SOC labels \cite{ref5,ref22}; (2) canary to a small region; (3) compare $\{\mathrm{FPR},\mathrm{P95}\}$ to SLOs; (4) promote globally with policy-as-code; (5) drift checks and model cards \cite{ref13}.

\subsubsection{Case Snapshots}
\textbf{Credential stuffing:} POST/login bursts, UA/TLS stability, low dwell $\rightarrow$ challenge at $s(\mathbf{x})>\tau$; shape to origin; blacklist high-confidence sources \cite{ref24,ref25,ref31}.\\
\textbf{Stealth L7 DDoS:} Elevated P99 RPS on expensive endpoints, cache-bypass, synchronized bursts $\rightarrow$ DRL policy raises per-ASN limits and injects puzzles; escalate to scrubbing on persistence \cite{ref7,ref24,ref28}.
\subsection{Datasets and Benchmarks}
While proprietary WAAP telemetry dominates production, open datasets remain valuable: UNSW-NB15 for flow features, CIC-IDS2017/2018 for modern traffic and app-layer behaviors, and Bot-IoT for IoT-scale floods \cite{ref65,ref66,ref67}. For bot detection, public sets are scarce; synthetic generation and privacy-preserving sharing are active needs. Evaluation should report ROC/PR curves, F1, P95/P99 latency overhead, cache HIT impact, and \emph{attack containment time} (ACT).
\subsection{Metrics and Evaluation}
Track: precision/recall/F1, ROC-AUC and PR-AUC; EER for bot classifiers \cite{ref9}. Operational SLOs: added P95/P99 latency (ms), PoP CPU\%, cache HIT impact, ACT $=t_{\text{block}}-t_{\text{onset}}$. Given $(\mathrm{TP},\mathrm{FP},\mathrm{FN},\mathrm{TN})$,
\begin{equation}
\mathrm{Precision}=\frac{\mathrm{TP}}{\mathrm{TP}+\mathrm{FP}}, \quad
\mathrm{Recall}=\frac{\mathrm{TP}}{\mathrm{TP}+\mathrm{FN}}.
\end{equation}
Tune $\tau$ to meet FPB while minimizing ACT.

\subsection{Platform Capabilities}
\begin{figure}[!t]
\centering
\includegraphics[width=\columnwidth]{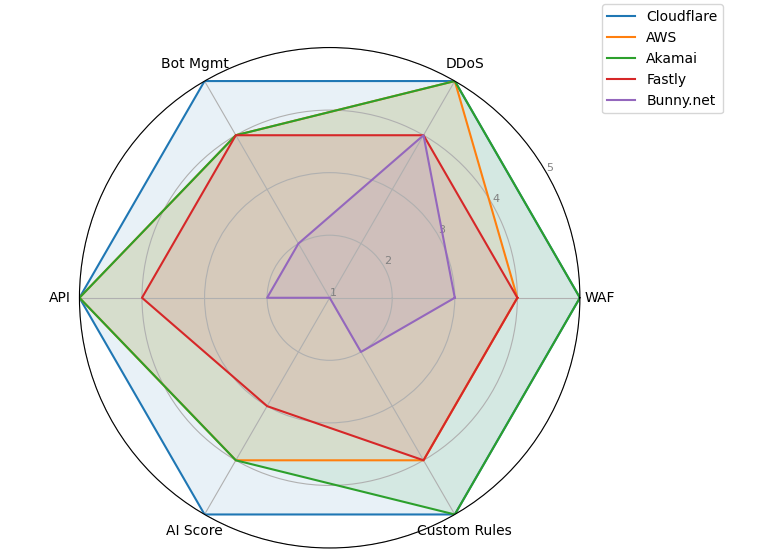}
\caption{WAAP providers across six AI-enabled dimensions}
\label{fig:radar}
\end{figure}

Table~\ref{tab:waap_large} summarizes WAAP capabilities across major CDNs \cite{ref24,ref25,ref27,ref28,ref30,ref31,ref32,ref34}. Figure~\ref{fig:radar} visualizes normalized scores.

\begin{table}[!t]
\caption{Normalized Scores for WAAP Feature Comparison }
\label{tab:radar_scores}
\centering
\footnotesize
\setlength{\tabcolsep}{4pt} 
\begin{tabular}{lccccc}
\toprule
Feature & Cloudflare & AWS & Akamai & Fastly & Bunny.net \\ \midrule
WAF         & 5 & 4 & 5 & 4 & 3 \\
DDoS        & 5 & 5 & 5 & 4 & 4 \\
Bot Mgmt    & 5 & 4 & 4 & 4 & 2 \\
API         & 5 & 5 & 5 & 4 & 2 \\
AI-feature coverage score   & 5 & 4 & 4 & 3 & 1 \\
Custom Rules& 5 & 4 & 5 & 4 & 2 \\
\bottomrule
\end{tabular}
\end{table}

\noindent\textbf{Interpretation:} Cloudflare and Akamai emphasize breadth of edge-native controls and learned models \cite{ref24,ref30,ref31}. AWS integrates deeply with Shield/WAF/Bot Control and IAM for policy coherence \cite{ref25,ref26,ref27}. Google couples Cloud Armor with load-balancing and sensitive-data tooling \cite{ref28,ref29}. Azure focuses on API gateway ``AI gateway'' and enterprise observability \cite{ref34}. Practical selection should weigh (i) policy-to-enforcement latency, (ii) regional model controls for compliance, (iii) first-class API discovery/positive modeling, and (iv) SIEM-friendly telemetry.

\subsection{Risk, Governance, and Compliance}
Adversarial ML risks (evasion and poisoning) can degrade detectors; countermeasures include adversarial training, robust statistics, and input sanitization \cite{ref55}. Data quality and XAI are central: skewed or sparse labels harm generalization, while faithful explanations support analyst trust and audits \cite{ref55}. Encrypted-traffic constraints mean DPI-free analytics must lean on side-channel metadata rather than payload inspection \cite{ref28,ref63}. Privacy and lawfulness require data minimization and retention controls; FL/DP/HE provide pathways toward GDPR and AI Act compliance \cite{ref19,sabbir,ref20}. Protocol hardening—preferring TLS~1.3 and QUIC—improves both security and performance \cite{ref64}. Finally, dual-use concerns persist: LLMs can aid both defenders and adversaries, so program controls and auditability are essential \cite{ref58}.

\noindent Governance primitives operationalize these principles: \emph{model cards} document data sources, slices, caveats, and intended use; \emph{DPIAs} record lawful bases, minimization, retention, and cross-border flows; \emph{red-team exercises} probe evasion and poisoning; and \emph{change control} gates model/rule pushes with audit trails. For behavioral features, ensure proportionality and opt-out paths where required \cite{ref13,ref20}. Encrypted-traffic analytics (TLS/QUIC side channels) avoid payload processing, reducing data-protection risk \cite{ref64,ref28}.


\section{Conclusion and Future Directions}
Edge-centric AI is now a cornerstone of web defense. CDNs provide the scale and proximity to convert telemetry into timely action, improving resilience against DDoS, bots, injection, and API abuse. Success depends on disciplined operations: privacy-preserving learning, robust MLOps, explainability, protocol hardening, and guardrails against adversarial pressure. With continued advances in XAI, robustness, and autonomous coordination, WAAP at the edge will further compress the window between attacker innovation and defender response. Limitations remain: evidence draws on heterogeneous datasets and a mix of academic and practitioner sources, and standardized WAAP benchmarks are still scarce.

Near-term priorities include a minimal WAAP-L7 benchmark; confidential edge inference via TEEs; federated learning with secure aggregation and differential privacy; privacy-preserving XAI with faithful, DP-bounded explanations; robust, byzantine-resilient federated learning with drift-aware personalization; leveraging 5G/6G edge fabrics and ledger-backed audit trails to harden telemetry integrity; active learning with uncertainty sampling and distribution-matched synthetic traffic to expand coverage where public data are scarce; and safety cases for security ML that tie metrics, tests, and guardrails to risk acceptance in production WAAP.

\section*{Acknowledgments}
We thank our colleagues and reviewers for their feedback. Large language models were used solely for editorial refinement; the authors remain fully responsible for all technical content and claims.

\bibliographystyle{IEEEtran}
\bibliography{references_cleaned}

\end{document}